\documentclass[review]{elsarticle}

\usepackage{amssymb, amsmath, amsthm, amsfonts, graphicx, dsfont, bbm,  bbold, url, color, mathrsfs, epsfig, subfig, multicol, tikz-cd,natbib,booktabs, lineno, hyperref, float}

\journal{Methods in Ecology and Evolution}

\begin{document}

\begin{frontmatter}
\title{ariaDNE: A Robustly Implemented Algorithm for Dirichlet Energy of the Normal}

\author{Shan Shan\fnref{myfootnote}}
\author{Shahar Z. Kovalsky\fnref{myfootnote}}
\author{Julie M. Winchester\fnref{biofootnote}}
\author{Doug M. Boyer\fnref{biofootnote}}
\author{Ingrid Daubechies\fnref{myfootnote}}
\fntext[myfootnote]{Department of Mathematics, Duke University}
\fntext[biofootnote]{Department of Evolutionary Anthropology, Duke University}

\begin{abstract}
Point 1: The geometric quantity {\it Dirichlet energy of the normal} measures how much a 3D surface bends; Dirichlet Normal Energy (DNE), a discrete approximation to that quantity for 3D mesh, is effective for morphological studies of anatomical surfaces. 

\begin{flushleft} Point 2: Recent studies found that DNE is sensitive to various procedures for preparing 3D mesh from raw scan data, raising concerns regarding comparability and objectivity when utilizing DNE in morphological research. We provide \underline{a} \underline{r}obustly \underline{i}mplemented \underline{a}lgorithm for computing the Dirichlet energy of the normal (ariaDNE) on 3D meshes. 
\end{flushleft}

\begin{flushleft}
Point 3: We show that ariaDNE is much more stable than DNE when preparation-related mesh surface attributes such as resolution (triangle count), mesh representation (i.e., a different set of points/nodes or triangles representing the same continuous surface) and noise are varied through simulation. We also show that the effects of smoothing and boundary triangles are more limited on ariaDNE than DNE. Further, ariaDNE retains the potential of DNE for biological studies, illustrated by it effectively differentiating species by dietary preferences.
\end{flushleft}

\begin{flushleft}
Point 4: ariaDNE can be a useful tool to uniformly quantify shape on samples of surface meshes collected with different instruments or at different resolutions, and prepared by varying procedures. To facilitate the field to move towards this goal, we provide scripts for computing ariaDNE and ariaDNE values for specimens used in previously published DNE analyses at \url{https://sshanshans.github.io/articles/ariadne.html}. 
\end{flushleft}

\end{abstract}

\begin{keyword}
DNE \sep morphology \sep shape characterization
\end{keyword}

\end{frontmatter}

\section{Introduction}
Developing methods to model and understand tempo and mode of macroevolution is an important goal for evolutionary biology (e.g., \citeauthor{harmon2010early}, \citeyear{harmon2010early}; \citeauthor{eastman2011novel}, \citeyear{eastman2011novel}; \citeauthor{revell2012phytools}, \citeyear{revell2012phytools}; \citeauthor{ingram2013surface}, \citeyear{ingram2013surface}) . Equally important are methods for effectively representing phenotypic differences between species (\citeauthor{adams2013geomorph}, \citeyear{adams2013geomorph}; \citeauthor{pampush2016introducing}, \citeyear{pampush2016introducing}; \citeauthor{winchester2016morphotester}, \citeyear{winchester2016morphotester}) without which many evolutionary modeling questions would be moot (\citeauthor{slater2018hierarchy}, \citeyear{slater2018hierarchy}). The potential for rapidly and objectively quantifying morphological phenotypes benefits greatly from the advent of easily accessible and widely available 3D digital models of anatomical structures. The unprecedented accessibility of 3D data is a direct result of technology improvements and cost reductions for generating them (\citeauthor{copes2016collection}, \citeyear{copes2016collection}), as well as proliferation and population of archives for sharing them (\citeauthor{boyer2016morphosource}, \citeyear{boyer2016morphosource}).

The new potential for better quantifications of shape is timely because of growing recognition that analyzing the wrong traits or poorly justified quantifications of traits may lead to mis-impressions about which processes meaningfully describe a clade's evolution (\citeauthor{slater2018hierarchy}, \citeyear{slater2018hierarchy}). Instead, it has been suggested that choice of morphological traits and the method for their quantification should be justified based on clade-specific hypotheses that propose not only an evolutionary mode but an ecological explanation.  In other words, demonstrating that one or more traits follow a particular evolutionary model does not go very far towards understanding the evolutionary processes at play in a clade, especially if there is no hypothesis relating variation in those traits to ecological variation. For instance, although \citeauthor{harmon2010early} (\citeyear{harmon2010early}) showed that the `adaptive radiation' (\citeauthor{osborn1902law},  \citeyear{osborn1902law}) or `early burst' (EB) model of evolution was rarely supported among dozens of clades tested, their study did not specify why the particular morphological traits they looked at should follow the EB model.  Showing that different traits can have different evolutionary patterns in the same clade, \citeauthor{meloro2010cats} (\citeyear{meloro2010cats}) found that tooth size and carnassial angle variables followed very different evolutionary patterns within Carnivora. Carnassial angle, arguably the more directly functional variable, followed an adaptive radiation model, while m1 size followed a more simple brownian motion model. As another example of the importance of trait function, \citeauthor{cantalapiedra2017decoupled} (\citeyear{cantalapiedra2017decoupled}) chose to quantify relative tooth crown height (hypsodonty) in order to understand drivers of disparity and diversity in equids, because hypsodonty has seemingly obvious adaptive significance for grazing in many clades, even beyond horses.  Moreover, hypsodonty has been formally demonstrated by \citeauthor{eronen2010impact} (\citeyear{eronen2010impact}) to be an ecometric (\citeauthor{eronen2010ecometrics}, \citeyear{eronen2010ecometrics}; \citeauthor{polly2015measuring}, \citeyear{polly2015measuring}) for grassland use in equids. 

A promising class of features are those that quantify the overall geometric quality of an object's  surface. They are referred to as ``shape characterizers'' and distinguished from ``shape descriptors'' (\citeauthor{evans2013shape}, \citeyear{evans2013shape}), the latter primarily including geometric morphometric quantifications of shape (\citeauthor{adams2013geomorph}, \citeyear{adams2013geomorph}).  Examples of shape characterizers include relief index (RFI) (e.g., \citeauthor{m2003occlusal}, \citeyear{m2003occlusal}), orientation patch count (OPC) (e.g., \citeauthor{evans2007high}, \citeyear{evans2007high}; \citeauthor{evans2014evolution}, \citeyear{evans2014evolution}; \citeauthor{melstrom2017relationship}, \citeyear{melstrom2017relationship}) and Dirichlet Normal Energy (DNE) (e.g., \citeauthor{bunn2011comparing}, \citeyear{bunn2011comparing}; \citeauthor{winchester2016morphotester}, \citeyear{winchester2016morphotester}; \citeauthor{pampush2016introducing},  \citeyear{pampush2016introducing}). RFI measures both the relative height and sharpness of an object; OPC measures the complexity or rugosity of a surface; DNE measures the bending energy of a surface. When different teeth exhibit substantially different values of a shape characterizer, they usually also look different and have easily conceivable functional and ecological differences. For instance, a tooth with higher relief often has sharper blades or longer, sharper cusps that cut or pierce food items more effectively than a tooth with lower relief. As another example, DNE differences among blood cells potentially correspond to the turbulence they induce in blood flow, or whether they tend to clog small arterioles.

DNE has several advantages compared to popular shape characterizers like RFI and OPC. First,  DNE is landmark-free and independent of the surface's initial position, orientation and scale making it less susceptible to observer induced error/noise. RFI and OPC rely on the orientation of the tooth relative to an arbitrarily defined occlusal plane. OPC also relies on the orientation of the tooth with regard to rotation around the central vertical axis. Second, direct comparisons show that DNE has a stronger dietary signal for teeth than RFI and OPC (\citeauthor{winchester2014dental}, \citeyear{winchester2014dental}). This greater success in dietary separation is likely due to its more effective isolation of information on the ``sharpness'' of surface features.  In contrast, RFI only measures the relative cusp and/or crown height which does not describe sharpness; OPC is less sensitive to changes in blade orientation due to its binning protocol (\citeauthor{boyer2010evidence}, \citeyear{boyer2010evidence}). 

DNE computes a discrete approximation to the {\it Dirichlet Energy of the normal}. This quantity is a mathematical attribute of a continuous surface, coming from differential geometry; it is defined as the integral, over the surface, of change in the normal direction, indicating at each point of the surface, how much the surface bends.  In practical applications, a continuous surface is represented as a triangular mesh, which can be described by a collection of points or nodes and triangles. (We note that the nomenclature is not standardized across all scientific fields; in computer science these would be called {\it vertices} and {\it triangular faces}, respectively; see e.g., \citeauthor{botsch2010polygon}, \citeyear{botsch2010polygon}). To compute DNE on such a discrete mesh, normal directions must be estimated for each point/triangle. The sum of the change of normal directions over the points/triangles is then used to approximate the Dirichlet energy of the normal for the continuous surface that the mesh represents.  However, the DNE algorithm published in {\it MorphoTester} (\citeauthor{winchester2016morphotester}, \citeyear{winchester2016morphotester}) and in the {\it R} package ``molaR'' (\citeauthor{pampush2016introducing} is sensitive to varying mesh preparation protocols and requires special treatment for boundary triangles, which are triangles that have one side/node that fall on the boundary of the mesh   (\citeauthor{pampush2016wear}, \citeyear{pampush2016wear}, \citeauthor{spradley2017smooth}, \citeyear{spradley2017smooth}), leading to concerns regarding the comparability and reproducibility when utilizing DNE for morphological research. 

\begin{figure}
 \begin{center}
    \includegraphics[width = 12cm]{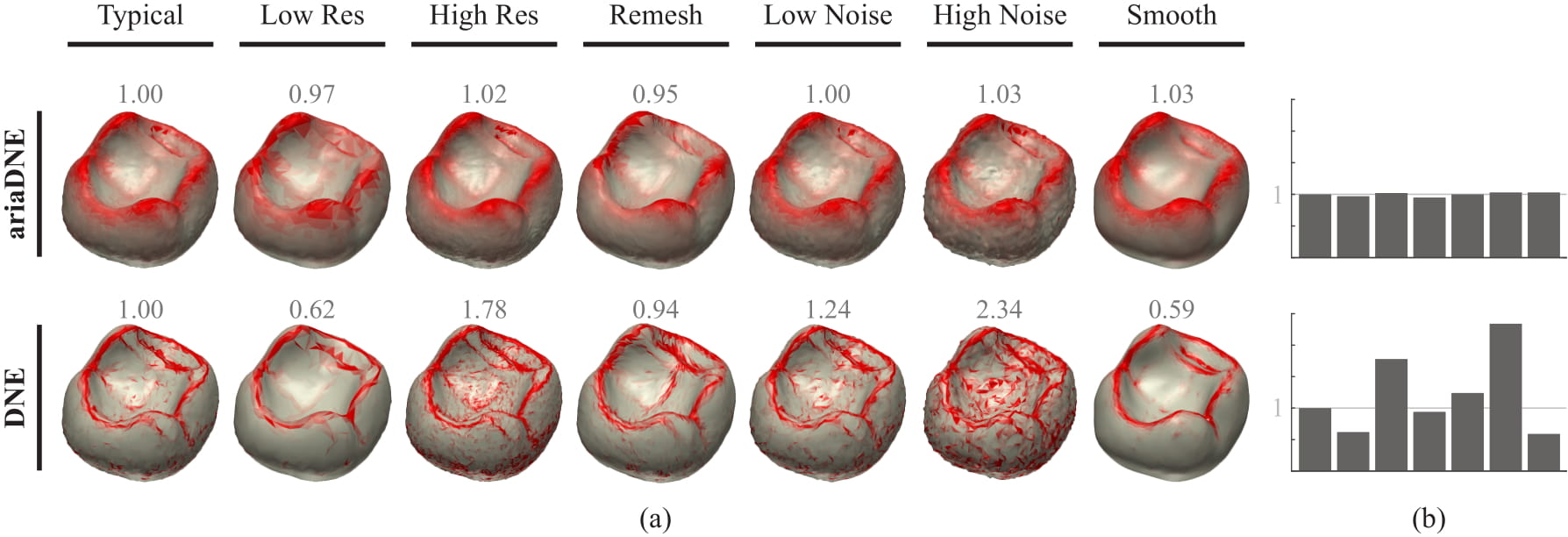}
     \end{center}
\caption{
Comparing effects of mesh resolution (triangle count), re-meshing, noise and smoothing on ariaDNE (top) and DNE (bottom). (a) shows the distribution of curvature as measured by each method overlaid in shades of red on a grey 3D rendering of the surface. ariaDNE and DNE values, normalized for comparability by the values for the typical tooth, are shown above each surface and summarized in the bar plots (b), demonstrating the robustness of ariaDNE versus DNE.
}
\label{fig:teaser}
\end{figure}

Recent attempts to address this issue have developed protocols for standardizing the mesh preparation process (\citeauthor{spradley2017smooth}, \citeyear{spradley2017smooth}). Unlike previous work, we provide \underline{a} \underline{r}obustly \underline{i}mplemented \underline{a}lgorithm for Dirichlet energy of the normal (ariaDNE), that is insensitive to a greater range of mesh preparation protocols. Fig.~\ref{fig:teaser} shows DNE and ariaDNE values on an example tooth. The red surface shading indicates the value of curvature as measured by each approach; it is uniformized across each row by the row's highest local curvature value. To demonstrate this insensitivity empirically, we test the stability of our algorithm on tooth models with differing triangle counts, remeshing/mesh representation (i.e., a different set of points/nodes or triangles representing the same continuous surface) and simulated noise. We also test the effects of smoothing and boundary triangles as in \citeauthor{spradley2017smooth} (\citeyear{spradley2017smooth}). We furthermore assess the dietary differentiation power of ariaDNE. 
 
\section{Materials and Methods}
\label{sec:method}

\subsection{Methodological analysis of ariaDNE}
\label{sec:ariadne}
\citeauthor{bunn2011comparing} (\citeyear{bunn2011comparing}) noted the relevance of the differential geometry concept of Dirichlet energy of the normal for morphology and provided an algorithm called DNE calculating an approximation to this quantity on discrete surface meshes by summing the local energy over all triangles. The local energy on a triangle is defined by the total change in the normals; this provides a local estimate for the curvature of the surface. However,  this change in normals is sensitive to how a continuous surface is discretized. That is, a different resolution (triangle count), mesh representation, or contamination by noise or small artifacts can all lead to significantly different numerical values. 

To address this sensitivity problem, we leverage the observation that the local energy can be also expressed by the curvature at the query point on the surface (\citeauthor{willmore1965note}, \citeyear{willmore1965note}); another simple method for estimating curvature on discrete surfaces is by Principal Component Analysis (PCA). The procedure is outlined as follows. For each query point, find all its neighboring points within a fixed radius; the value of this radius is set as a parameter for the method (\citeauthor{yang2006robust}, \citeyear{yang2006robust}). Then apply PCA to the coordinates of those points; the plane spanned by the first two principal components typically approximates the tangent plane to the surface at the query point, with the third principal component approximating the normal direction. The corresponding smallest principal component score
$\sigma = \lambda_0/ (\lambda_0 + \lambda_1 + \lambda_2), ~\mbox{where} ~\lambda_0<\lambda_1<\lambda_2.$
indicates the deviation from the fitted plane, i.e. the curvature. 

\begin{figure}
   \begin{center}
    \includegraphics[width = 5cm]{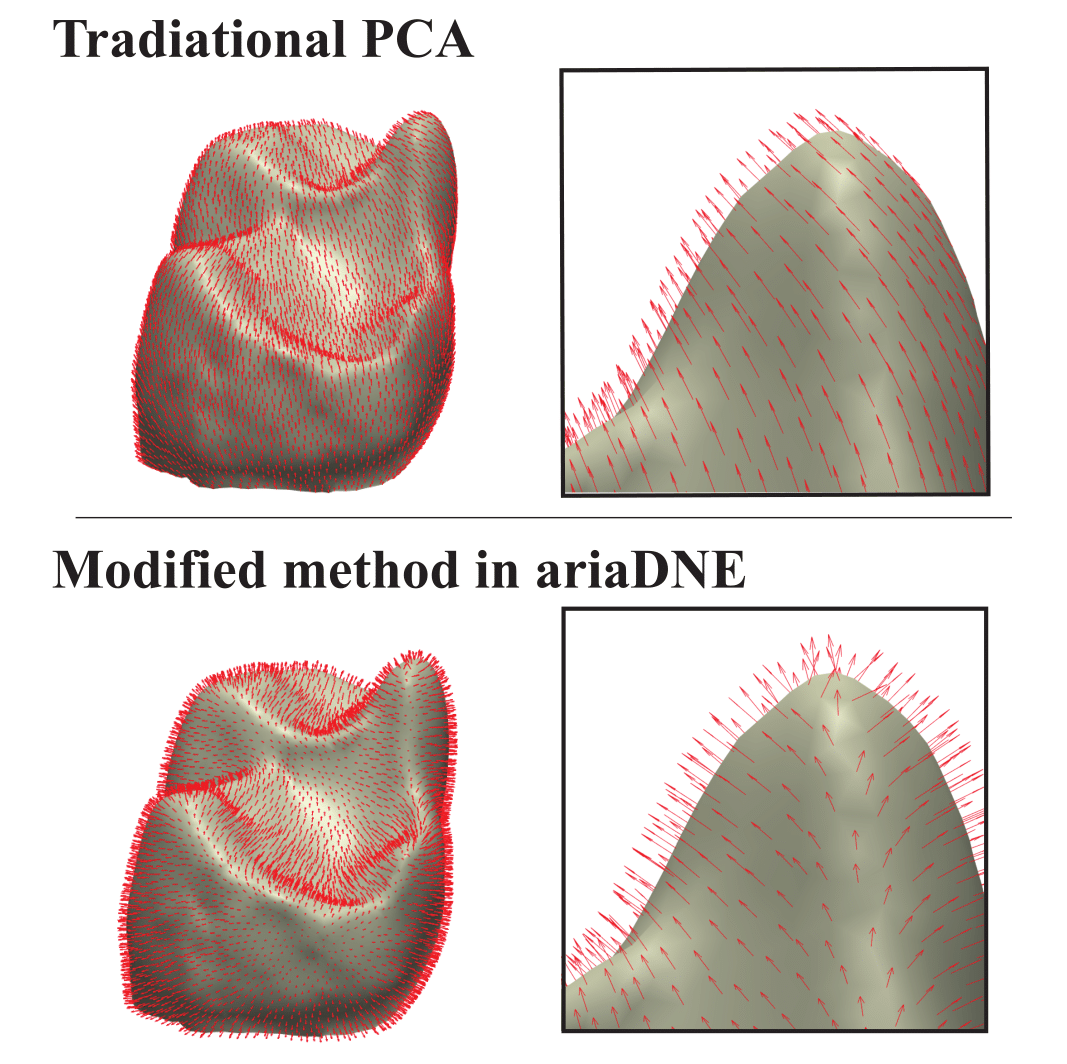}
     \end{center}
    \caption{Improved normal estimation with our modified PCA method. Top: traditional PCA method gives skewed normal estimates on a pointed cusp, leading to erroneous curvature approximation. Bottom: our method gives better normal approximation, and therefore improves curvature approximation.}
    \label{fig:normals}
\end{figure}

There are two issues with this PCA method: (1) The third principal component does not always approximate the surface normal, and therefore the smallest principal component score may not accurately reflect curviness as we discussed above, that is, the deviation of the surface from the tangent plane. Fig.~\ref{fig:normals} (top) shows an erroneous normal approximation for a pointed cusp, where the normals should be perpendicular to the surface but using standard PCA gives skewed estimation. (2) Standard PCA becomes numerically unstable (due to ill-conditioning) when the number of nearby neighbors is low. This implies that when the mesh is of low resolution, there may not be enough points to conduct PCA. 

To resolve the first issue, we modify the algorithm to choose at each query point the principal component closest to its normal, and set the curvature at that point to be the score of the chosen principal component. Fig.~\ref{fig:normals} (bottom) illustrates the effects of this simple modification, which produces estimates more consistent with surface normals, thereby providing a better local estimate of the tangent plane, and in turn curvature. In practice, normals at a point are obtained by taking a weighted average of normals of adjacent triangles, easily computed on discrete meshes. 

To resolve the second issue, we propose a modification to the traditional PCA method. Selecting the neighbors within a fixed radius could result, near some point, in a small-sized neighborhood where few or even no points would be selected; instead, we apply a ``weighted PCA'', with weights decaying according to the distance away from the query point, retaining the rest of procedure. There are many ways to define the weight function. The traditional PCA method chooses the weight function to be the indicator function over the set of points within an a priori specified distance from the query point (i.e. the weight is one for the points within a fixed radius and zero elsewhere). For ariaDNE, we set the weight function to be the widely-used Gaussian kernel $f(x) = e^{-x^2/ \epsilon^2}$. 

\begin{figure}
   \begin{center}
    \includegraphics[width = 12cm]{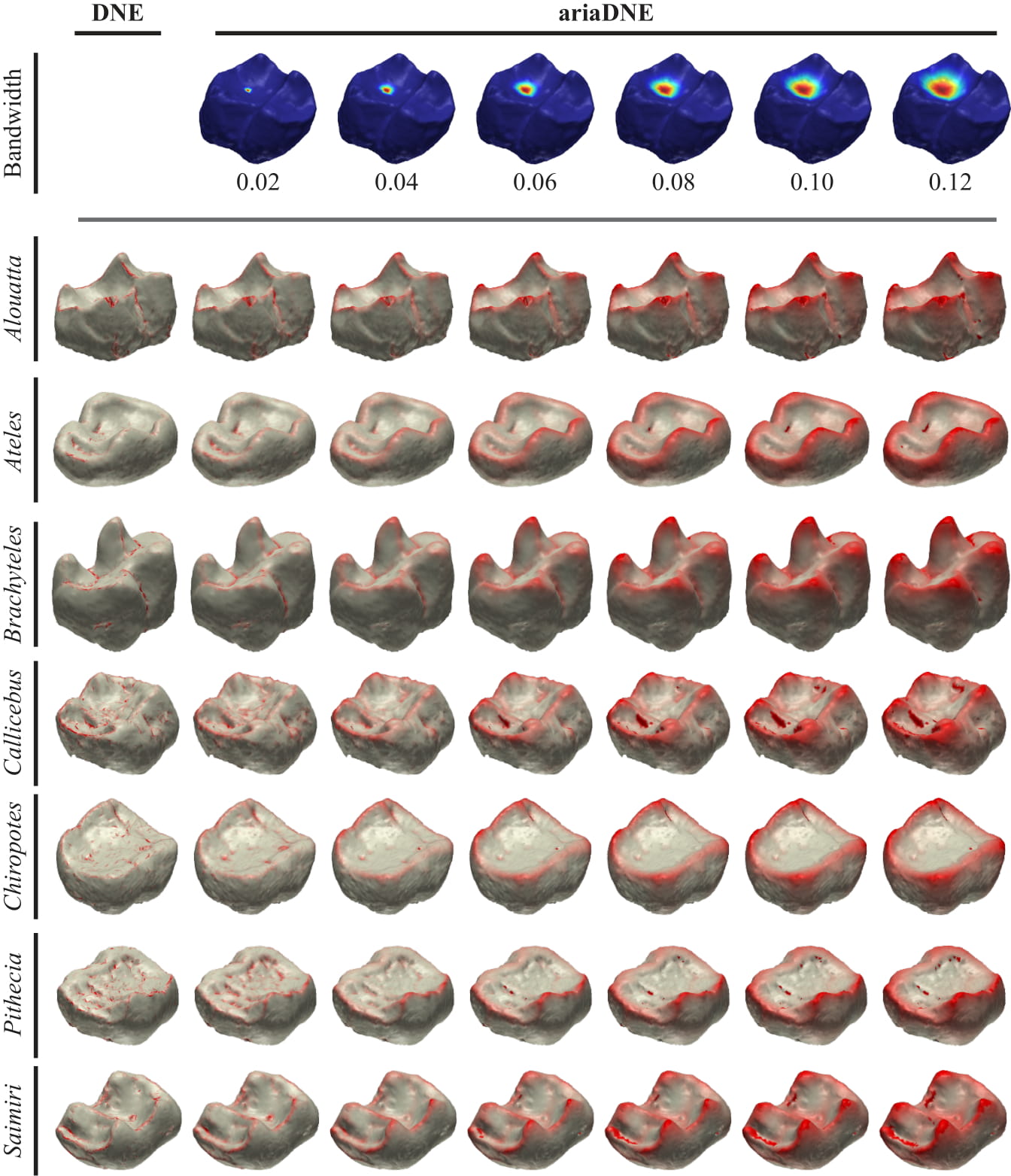}
        \end{center}
    \caption{Effect of increasing the $\epsilon$ parameter (bandwidth) on the weight function (top; red indicates highest weight) and curvature computed by ariaDNE for molar teeth {\it Alouatta}, {\it Ateles}, {\it Brachyteles}, {\it Callicebus}, {\it Chiropotes}, {\it Pithecia}, {\it Saimiri}. Choices for $\epsilon$ are 0.02, 0.04, 0.06, 0.08, 0.10, 0.12, witch surface shading similar to Figure~\ref{fig:teaser}. When $\epsilon$ is small, both DNE and ariaDNE capture fine-scale features on the tooth. When $\epsilon$ is larger, ariaDNE captures larger scale features.} 
        \label{fig:plate}
\end{figure}

The Gaussian kernel captures local geometric information on the surface. The parameter $\epsilon$ indicates the size of local influence. Figure~\ref{fig:plate} illustrates effects of different $\epsilon$ on the weight function: the larger $\epsilon$, the more points on the mesh have significant weight values, resulting in larger principal component scores for those points. In consequence, when $\epsilon$ increases, local energy for each point becomes larger, and therefore ariaDNE becomes larger. In practice, we suggest using $\epsilon$ ranging from 0.04 to 0.1.  If $\epsilon$ is too small, the computed ariaDNE score will be highly sensitive to trivial features of the surface that are most likely to be noise (similar to traditional DNE); If $\epsilon$ is too large, the approximation will simply become non-local. Choosing an appropriate value of $\epsilon$ depends on the application in hand. 

In summary, we apply a weighted PCA, localized around each query point by means of the Gaussian kernel function. Then we find the principal component that is closest to its normal and set the curvature to be its principal score. AriaDNE is then computed by integrating this curvature estimate along the surface. For the exact procedure, see Appendix A in the supplementary materials. 

\subsection{Study samples}
Understanding the correlation between surface geometry and a metric like DNE or ariaDNE helps understand whether these metrics are relevant to questions concerning morphology, ecology and evolution. The meaningfulness and success of a metric have to be measured against relevant samples and the research questions. 

Here we use a sample of new world monkey (platyrrhine) second mandibular molars downloaded from {\it Morphosource} (\citeauthor{winchester2014dental}, \citeyear{winchester2014dental}). The sample has significant inter-specific breadth (7 genera)  and intra-specific depth (10 individuals per genus). It consists of meshes (117,623 - 665,001 points, 234,358 - 1,334,141 triangles) from 7 extant platyrrhine primate genera: {\it Alouatta}, {\it Ateles}, {\it Brachyteles}, {\it Callicebus}, {\it Chiropotes}, {\it Pithecia}, and {\it Saimiri}. Platyrrhine dentitions have been essential for questions about dental variation and dietary preference (\citeauthor{anthony1993tooth}, \citeyear{anthony1993tooth}; \citeauthor{dennis2004dental}, \citeyear{dennis2004dental};  \citeauthor{ledogar2013diet}, \citeyear{ledogar2013diet}; \citeauthor{winchester2014dental}, \citeyear{winchester2014dental}; \citeauthor{allen2015dietary}, \citeyear{allen2015dietary}; \citeauthor{pampush2016wear} 2016 a,b). Questions have included how dietarily diverse platyrrhines should be considered based on available behavioral data, whether and how dental morphology is reflective of diet differences, and how important tooth wear, individual variation, and scale of geometric features are when considering tooth differences between species. In the following sections, we tested the stability of ariaDNE by perturbing attributes like triangle count and mesh representation. We also tested the effects of noise, smoothing and boundary triangles on ariaDNE. Furthermore, we assessed its power in differentiating the 7 platyrrhine primate species according to dietary habits.    

\subsection{Sensitivity test}
\subsubsection{Triangle count}
\label{sec:one}
To evaluate the sensitivity of ariaDNE under varying mesh resolution, each tooth was downsampled to produce simplified surfaces with  20k, 16k, 12k, 8k, 4k, and 2k triangles. We computed ariaDNE values ($\epsilon = 0.04, 0.06, 0.08, 0.1$ using the MATLAB function ``ariaDNE'' provided in Section~\ref{sec:code}. For comparison, we also computed traditional DNE values using the function ``DNE''  (Section~\ref{sec:code}), a MATLAB port of the R function  ``DNE''  from ``molaR".  Default parameters were used for DNE, with outlier percentile at 0.1 and boundary triangles excluded. 

\subsubsection{Mesh representation}
\label{sec:two}
\begin{figure}[H]
   \begin{center}
   \includegraphics[width = 10cm]{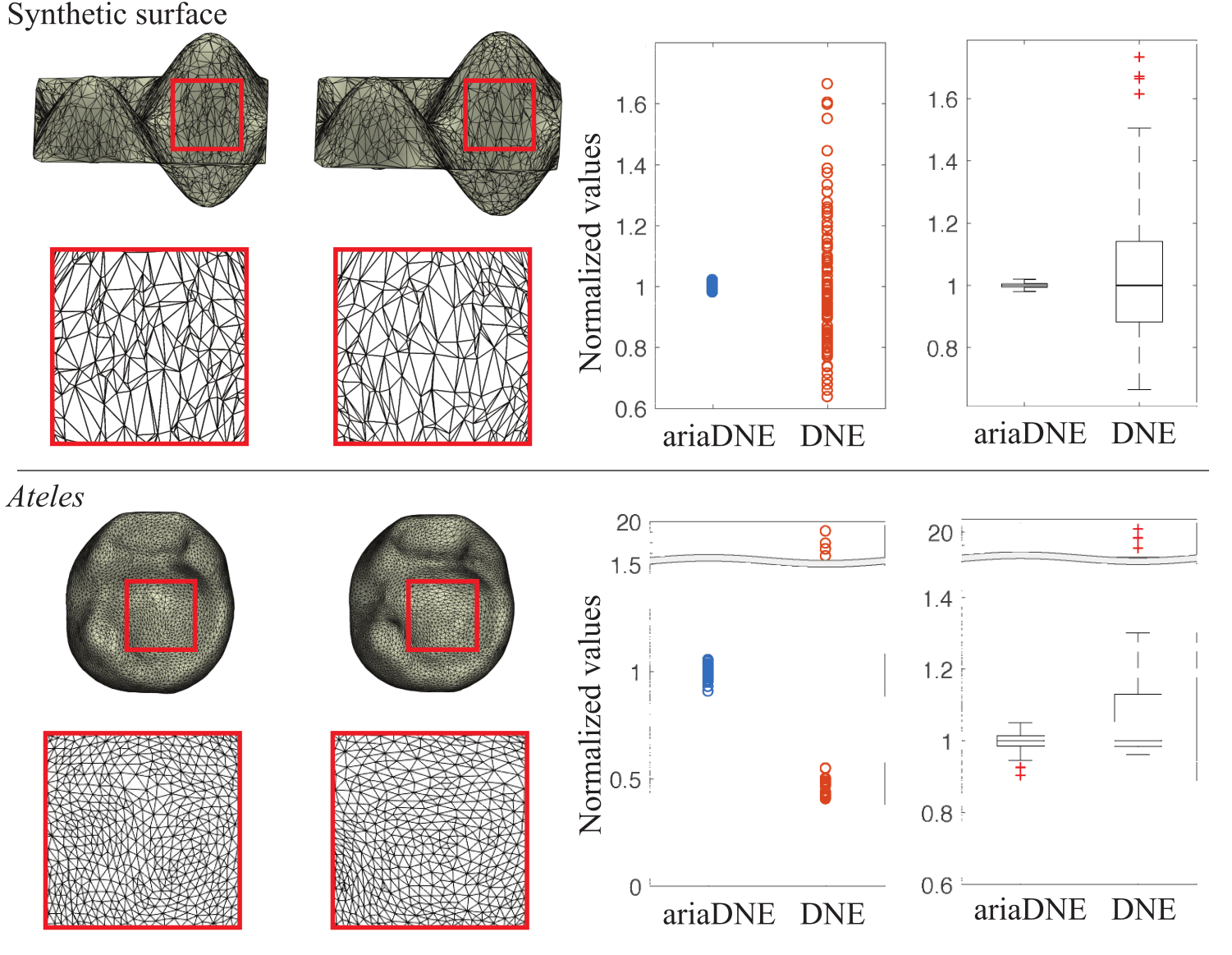} 
    \end{center}
    \caption{Effect of varying mesh representations on ariaDNE and DNE values computed for a synthetic surface (top) and a tooth from {\it Ateles} (bottom). Left panel: examples of different mesh representations. Right panel: scatter plots and box plots of ariaDNE ($\epsilon = 0.08$) and DNE values computed for $N$ meshes representing the synthetic surface (top, $N = 100$) and the tooth surface (bottom, $N = 70$).}
    \label{fig:triangulation}
\end{figure}

A continuous surface can be represented by different discrete meshes; even with the same resolution (triangle count), they can differ by altering the position of points and their adjacency relations (i.e., triangles). We would like ariaDNE to be roughly the same for all meshes that represent the same continuous surface. To evaluate the sensitivity of ariaDNE under varying mesh representations, we tested on a surface generated by a mathematical function as well as real tooth samples. First, we tested it on the surface $S$ defined by $z = 0.3\sin(2x)\sin(2y)$ where $0 \leq x \leq 1$ and $0 \leq y \leq 1$ (Fig.~\ref{fig:triangulation}). To generate a mesh for this explicitly defined surface, we randomly picked 2000 sets of $(x,y)$ coordinates uniformly distributed on $0 \leq x \leq 1$ and $0 \leq y \leq 1$ and calculated their accompanying $z$-values using the equation above. Each set of $(x,y,z)$ coordinates represented a node/point in the mesh, and the triangles are obtained by applying Delaunay Triangulation to these points. We generated 100 meshes by repeating these steps and computed their DNE and ariaDNE values as in \ref{sec:one}. We remark here that meshes generated by this procedure do not necessarily have evenly distributed points; some areas of the mesh can have finer resolution than others.

Real tooth samples are already given as meshes; we generated new mesh representations for each tooth sample by computing pairwise surface correspondences. Specifically, points and their adjacency relations from one surface were taken to the other surface in the samples by correspondence maps computed using the methods 
in (\citeauthor{boyer2011algorithms}, \citeyear{boyer2011algorithms}), between all pairs of surfaces in the sample. These correspondences resulted in 70 different mesh representations for each tooth in the sample. We computed their DNE and ariaDNE ($\epsilon = 0.04, 0.06, 0.08, 0.1$) as in \ref{sec:one}. 

\subsubsection {Simulated noise}
\begin{figure}
   \begin{center}
   \includegraphics[width = 10cm]{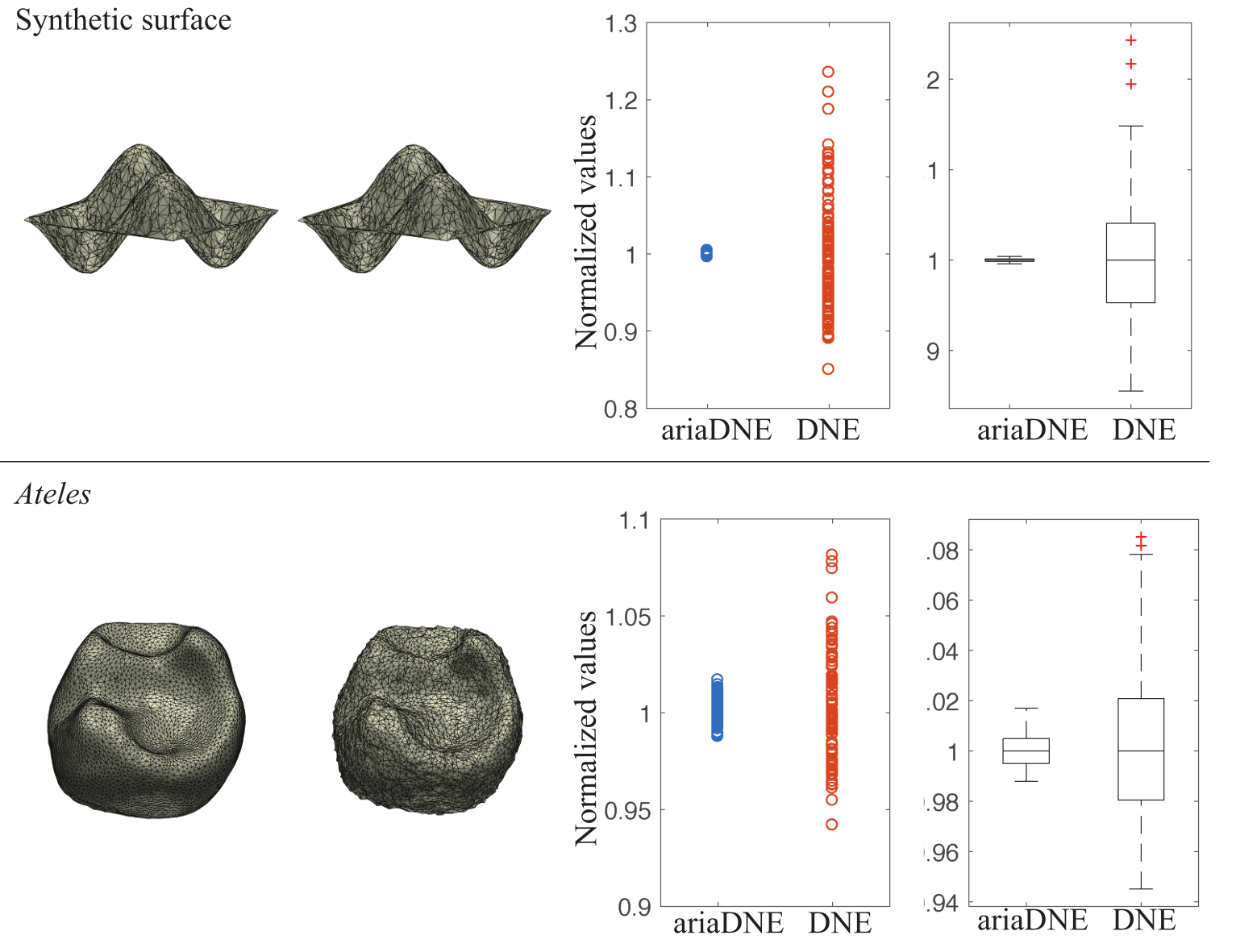} 
    \end{center}
    \caption{Effect of simulated noise on ariaDNE and DNE values computed for a synthetic surface (top) and a tooth from {\it Ateles} (bottom). Left panel: examples of original surfaces (left) and noisy surfaces (middle). Right panel: 
scatter plots and box plots of ariaDNE ($\epsilon = 0.08$) and DNE values computed for 100 meshes with random noise representing the synthetic surface (top) and tooth surface (bottom).}
    \label{fig:noise}
\end{figure}

To evaluate the sensitivity of ariaDNE to small artifacts on the surface, we tested it when simulated noise was added to the surface defined as in \ref{sec:two} as well as real tooth samples. First, given a mesh representing the same surface $S$ as in \ref{sec:two}, a noisy mesh was obtained by adding a random variable uniformly distributed on $[-0.001, 0.001]$ to the $x, y, z$ coordinates of each node/point on the mesh (Fig.~\ref{fig:noise}). We then generated 100 noisy versions of the given mesh by repeating the previous steps. 

For real tooth data, we generated a noisy mesh by adding a random variable uniformly distributed on $[-0.003, 0.003]$ to the $x, y, z$ coordinates for each node/point in the mesh (Fig.~\ref{fig:noise}). The noise level was chosen arbitrarily; we added more noise to the tooth samples to increase diversity of the test cases. We obtained 100 noisy meshes per tooth, and computed their DNE and ariaDNE (with $\epsilon = 0.04, 0.06, 0.08, 0.1$) values as in \ref{sec:one}.

\begin{figure}[H]
   \begin{center}
   \includegraphics[width = 12cm]{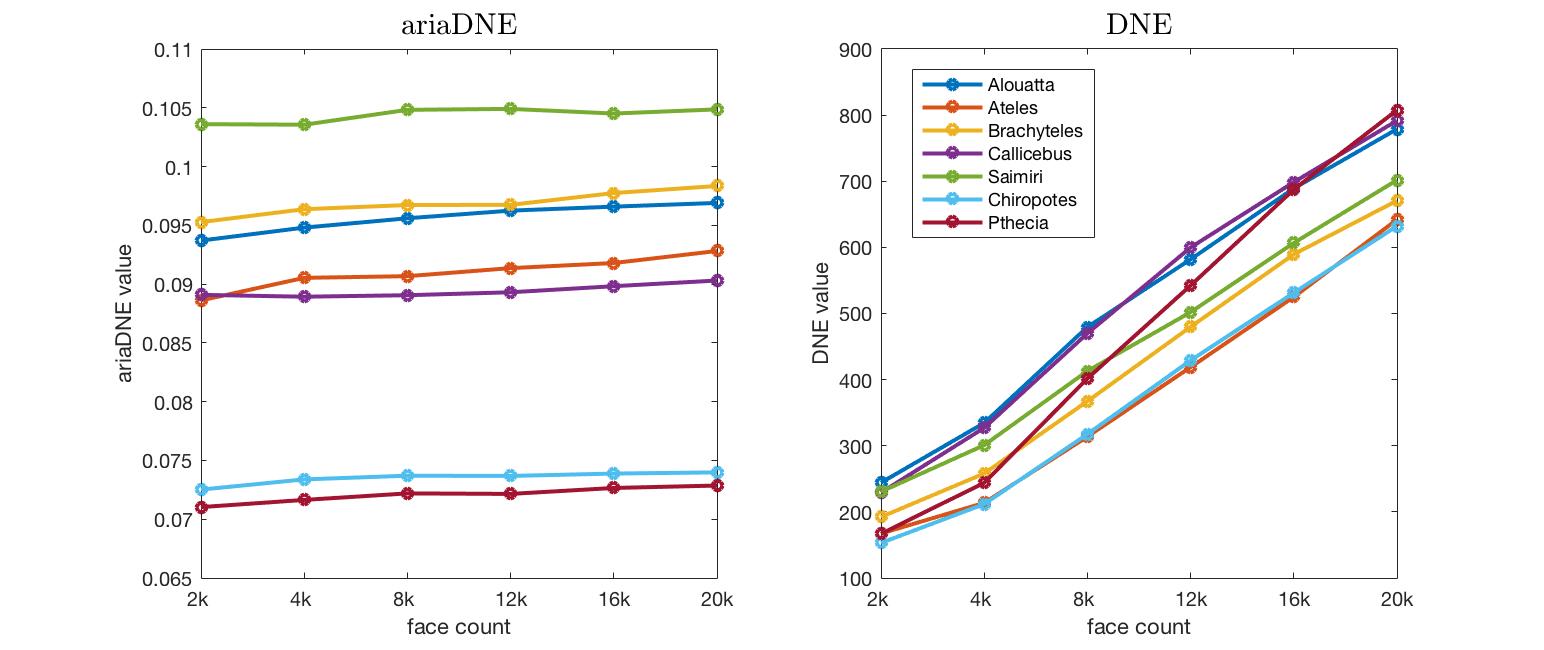} 
    \end{center}
    \caption{Effect of increasing triangle count on ariaDNE (left) and DNE (right) values computed for 7 teeth from {\it Alouatta}, {\it Ateles}, {\it Brachyteles}, {\it Callicebus}, {\it Saimiri}, {\it Chiropotes}, {\it Pithecia}. The ariaDNE values for each tooth remain relatively unchanged, compared to the DNE values, under varying resolution/triangle counts.}
    \label{fig:face}
\end{figure}

\subsubsection{Smoothing}
Smoothing is commonly used to eliminate noise produced during scanning, segmentation, and reconstruction. \citeauthor{spradley2017smooth} (\citeyear{spradley2017smooth}) tested the effects of various smoothing operators and smoothing amounts on DNE with surface meshes of hemispheres and primate molars. They suggested that aggressive smoothing procedures like Laplacian smoothing and implicit fairing should be avoided. To evaluate the performance of our method under different smoothing algorithms, we randomly picked 7 tooth models from our sample (one from each taxa) and generated their smooth surfaces by applying 100 iterations of the {\it Avizo} smoothing module, 3 iterations of the {\it Meshlab} function {\it HC Laplacian Smoothing}, or 3 iterations of the {\it implicit fairing method} using {\it MorphoTester}. Then we computed their DNE and ariaDNE ($\epsilon = 0.08$) as in \ref{sec:one}. 

To further evaluate the performance of our method under varying amounts of Avizo smoothing, we iteratively applied the {Avizo} smoothing module to a single molar tooth from {\it Ateles}. The smoothing function was performed in intervals of 20 on the raw surface mesh, evenly spaced from 20 to 200 to generate 10 new surface meshes. Default value for lambda was kept (lambda = 0.6). We then computed their DNE and ariaDNE ($\epsilon = 0.08$) as in \ref{sec:one}. 

\subsubsection{Boundary triangles}
Triangles with one side/node that are on the boundary of the mesh have a large impact on traditional DNE, calling for special treatment (\citeauthor{spradley2017smooth}, \citeyear{spradley2017smooth}). We assess how such boundary triangles affect ariaDNE on two molar teeth, 
one of {\it Ateles} where crown side walls are relatively bulged outwardly, and one from  
{\it Brachyteles} where crown side walls are relatively unbulged (Fig.~\ref{fig:boundary}). For each tooth, we found its boundary triangles and computed their local energy using both ariaDNE and DNE (``BoundaryDiscard"  = ``none'', i.e., no boundary triangles will be removed). 

\subsection{Tests on species differentiation}
\label{sec:three}
Previous studies revealed systematic variation among species with different dietary habits in DNE values and other topographic metrics, such as RFI.  To test the differentiation power for species and their dietary preferences, we compared RFI, DNE and ariaDNE on the 70 lower second molars in our sample that belong to 7 taxa, {\it Alouatta}, {\it Ateles}, {\it Brachyteles}, {\it Callicebus}, {\it Chiropotes}, {\it Pithecia}, and {\it Saimiri}. The diet-classification scheme from \citeauthor{winchester2014dental} (\citeyear{winchester2014dental}) was used to identify {\it Alouatta} and {\it Brachyteles} as folivorous,  {\it Ateles} and {\it Callicebus} as frugivorous,  {\it Chiropotes} and {\it Pithecia} as hard-object feeding, and {\it Saimiri} as insectivorous. For each tooth we computed its RFI, DNE and ariaDNE ($\epsilon = 0.02, 0.04, 0.06, 0.08, 0.1, 0.12$) as in \ref{sec:one}. We then used ANOVA and multiple comparison tests to assess their differentiation power for dietary preferences. 

\section{Results}
\subsection{Sensitivity tests}
In numerical analysis, an algorithm is stable if perturbing inputs do not significantly affect outputs. To enable comparison, the change in the outputs can be quantified by {\it coefficient of variation}, which is the ratio of the standard deviation to the mean. We perturbed each mesh in the sample by varying the number of triangles, changing the mesh representation or adding simulated noise. Supplementary Tables 1-4 provide coefficients of variation of the DNE and ariaDNE values of the perturbed meshes in each collection per tooth model. For each tooth and each perturbed collection, the coefficient of variation of ariaDNE is less than that of DNE, meaning ariaDNE is relatively more stable than DNE under varying resolution/triangle count, remeshing/mesh representation and noise. Table~\ref{table:all} summarizes results, indicating means of coefficients of variation from Supplementary Tables 1-4.  Fig.~\ref{fig:face} illustrates effects of increasing triangle count on ariaDNE ($\epsilon = 0.10$) and DNE values computed for 7 arbitrarily chosen teeth from {\it Alouatta}, {\it Ateles}, {\it Brachyteles}, {\it Callicebus}, {\it Saimiri}, {\it Chiropotes}, {\it Pithecia}. The ariaDNE values for each tooth (maximum percent change: 3.42 \%) remain relatively unchanged under varying resolution/triangle counts, compared to the DNE values (maximum percent change: 384\%).  Figs.~\ref{fig:triangulation} and \ref{fig:noise} compare normalized DNE and ariaDNE values computed for a synthetic surface and a tooth by varying the mesh representation and adding noise. In the scatter plots, the ariaDNE and DNE values are normalized to have a mean one in each case; in the box plots, the values are normalized to have a median one in each case. 

 \begin{table}
\centering
\resizebox{0.5\textwidth}{!}{%
\begin{tabular}{llrrr}
\toprule
\multicolumn{2}{c}{Method} & Triangle Count & Remeshing & Noise\\
\midrule
ARIADNE & $\epsilon = 0.04$ & 0.0213 & 0.0824 & 0.0055 \\
 & $\epsilon = 0.06$ & 0.0114 & 0.0429 & 0.0044 \\
 & $\epsilon = 0.08$ & 0.0117 & 0.0304 & 0.0039 \\
 & $\epsilon = 0.10$ & 0.0117 &0.0293 & 0.0038 \\
DNE & & 0.420 & 2.3075   &0.0169\\
\bottomrule
\end{tabular}}
\caption{Robustness of ariaDNE under various mesh attributes perturbation. For each tooth in the 70 platyrrhine sample, we generated three collections of perturbed meshes by varying the number of triangles, changing the mesh representation or adding simulated noise. We computed the coefficient of variation of their DNE and ariaDNE values in each collection for each tooth (see supplementary tables 1-4). The numbers in the table are obtained by taking the mean across all 70 tooth samples.}
\label{table:all}
\end{table}

Table~\ref{table:smooth} shows percent change of ariaDNE and DNE values subject to different smoothing algorithms. After 100 iterations of {Avizo} smoothing, ariaDNE increased 2\% of its original value whereas traditional DNE dropped to 46\%. After 3 iterations of HC Laplacian smoothing and implicit fairing, ariaDNE dropped to approximately 90\% of the original value whereas traditional DNE dropped to approximately 40\%. The larger drop in values using Laplacian smoothing and implicit fairing is consistent with the discussion by \citeauthor{spradley2017smooth} (\citeyear{spradley2017smooth}). However, for all smoothing algorithms, the variation in ariaDNE is significantly lower than for traditional DNE. This suggests that ariaDNE is relatively stable under varying smoothing algorithms. 

\begin{table}
\centering
\resizebox{\textwidth}{!}{%
\begin{tabular}{lrrrrrcrrrrr}
\toprule
& \multicolumn{5}{c}{DNE} & \phantom{a}& \multicolumn{5}{c}{ariaDNE} \\
\cmidrule{2-6} \cmidrule{8-12}
& Raw & Avizo & Laplacian & Fairing &COV && Raw & Avizo & Laplacian & Fairing &COV \\
\midrule
{\it Alouatta} & 1 & 0.38 & 0.35 & 0.48 &0.120  &&1 & 1.00 & 0.90 & 0.90 &0.049\\
{\it Ateles} & 1 & 0.57 & 0.46 & 0.57 &0.144 &&1 & 1.05 & 0.93 & 0.93 & 0.056 \\
{\it Brachyteles} & 1 & 0.54 & 0.45 & 0.57 &0.127 && 1 & 1.06 & 0.96 & 0.95 & 0.053\\
{\it Callicebus} & 1 & 0.46 & 0.39 & 0.52 &0.122 && 1& 0.99 & 0.90 & 0.91 &0.074 \\
{\it Chiropotes} & 1 & 0.51 & 0.37 & 0.48 &0.112 && 1 & 1.02 & 0.95 & 0.94 &0.06  \\
{\it Pithecia} & 1 & 0.43 & 0.29 & 0.41 &0.165  && 1 & 1.02 & 0.94 & 0.93 &0.058 \\
{\it Saimiri} & 1 & 0.33 & 0.44 & 0.56 &0.169  &&1 & 1.00 & 0.90 & 0.92 &0.056 \\
\bottomrule
\end{tabular}}
\caption{Effect of different smoothing algorithms on DNE and ariaDNE ($\epsilon = 0.08$) computed on the surfaces of molar teeth from {\it Alouatta}, {\it Ateles}, {\it Brachyteles}, {\it Callicebus}, {\it Chiropotes}, {\it Pithecia}, {\it Saimiri}. The numbers in the table are DNE and ariaDNE values divided by values of raw surfaces indicating the percent change. The table also contains coefficients of variation (COV) of DNE and ariaDNE computed on the three smooth surfaces in each taxa. The table demonstrates: (1) The effect of smoothing is limited on ariaDNE versus DNE. (2) ariaDNE is relatively more stable under varying smoothing algorithms.}
\label{table:smooth}
\end{table}

For Avizo smoothing, results show that both DNE and ariaDNE decrease for the first 40 iterations of smoothing. After approximately 40 iterations, ariaDNE increases and the traditional DNE continues to decrease until 100 iterations. After 100 iterations both start to increase. Perhaps most importantly, the degree of overall change in ariaDNE from unsmoothed surfaces to smoothed surfaces is much less than the overall change in traditional DNE. The increase on DNE values with smoothing is caused by mesh artifacts created during smoothing. After 40 iterations of smoothing, cusps of the smoothed mesh grow taller, while the basin becomes lower. Overall, ariaDNE have a smaller variance than the traditional implementation. This suggests that ariaDNE is relatively more stable under varying Avizo smoothing iterations.

\begin{figure}[H]
   \begin{center}
    \includegraphics[width = 5cm]{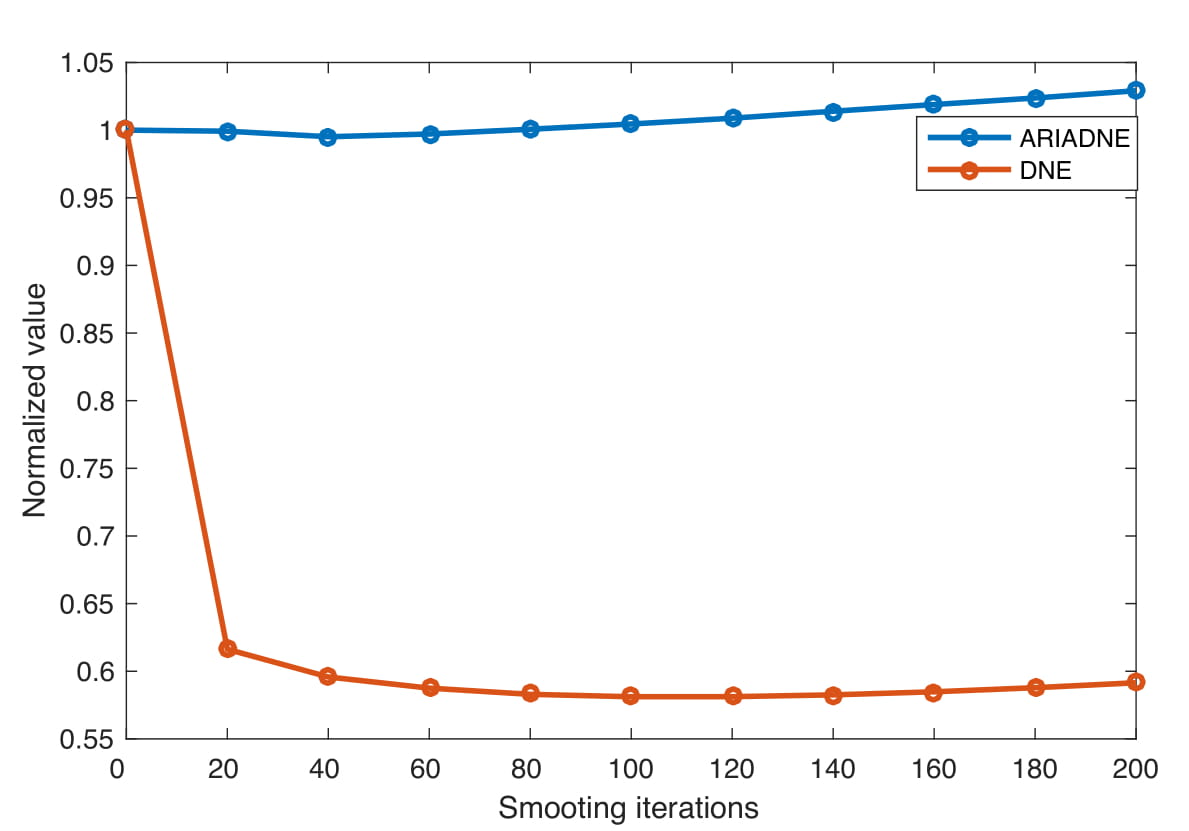} 
    \end{center}
    \caption{Percent change of varying {\it Avizo} smoothing amounts on ariaDNE ($\epsilon  = 0.08$) and DNE values computed for an {\it Ateles} molar tooth. }
    \label{fig:smoothing}
\end{figure}

\begin{figure}
   \begin{center}
    \includegraphics[width = 6cm]{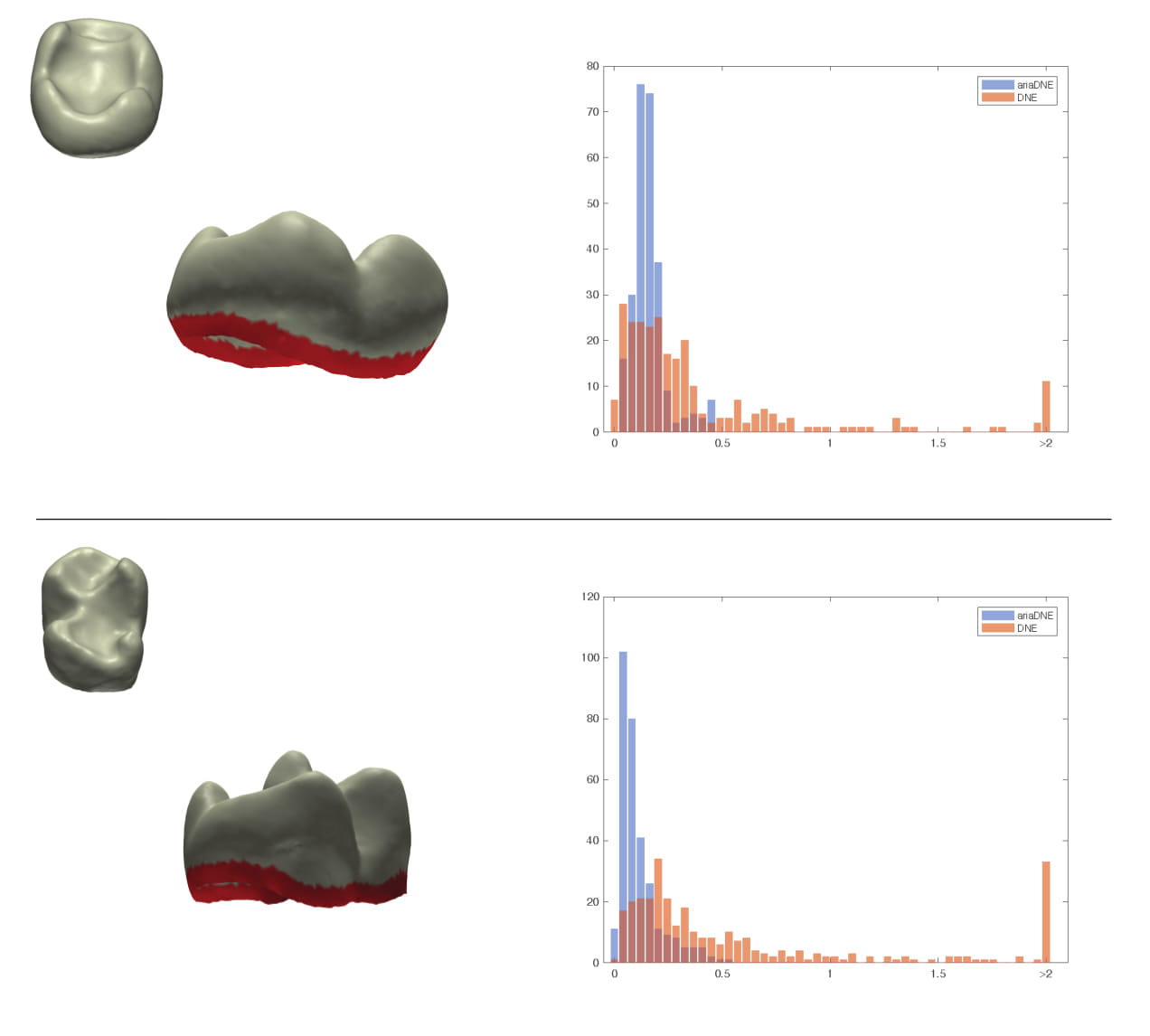} 
    \end{center}
    \caption{The boundary triangles have less impact on ariaDNE than DNE. The left panel shows an {\it Ateles} molar (top), with  a curvy side wall and a {\it Brachyteles} molar (bottom), with a straight side wall. The right panel shows histograms of local energy values of the boundary triangles, computed by ariaDNE and DNE. To enable comparison, the values are normalized by the mean of those of all triangles.}
 \label{fig:boundary}
\end{figure}

Fig.~\ref{fig:boundary} shows that the local energy of the boundary triangles computed with ariaDNE are among the smallest, whereas those computed with DNE have a few larger ones, which affect the DNE value for the whole surface. This histogram suggests that the effects of boundary triangles on ariaDNE are limited, and therefore no special treatment for them is needed. This represents another improvement for ariaDNE compared to DNE. 

\subsection{Species differentiation power}
\label{sec:four}

\begin{table}
\centering
\resizebox{0.5\textwidth}{!}{%
\begin{tabular}{lrrrrrrrr}
\toprule
&RFI & DNE &  \multicolumn{6}{c}{ariaDNE} \\
	&	& & $\epsilon = 0.02$ & 0.04 &0.06 &0.08 &0.10 &0.12 \\
\midrule
Fo-Fr   &0.0001    &0.0224    &0.0512    &0.8362    &0.0207    &0.0002    &0.0000    &0.0000 \\
Fo-H    &0.0000    &0.0000    &0.0010    &0.0003    &0.0000    &0.0000    &0.0000    &0.0000\\
Fo-I    &0.2414    &0.8463    &0.8986    &0.7564    &0.6544    &0.5584    &0.5893    &0.7376\\
Fr-H    &0.9372    &0.2295    &0.5425    &0.0049    &0.0000    &0.0000    &0.0000    &0.0000\\
Fr-I    &0.1888    &0.3910    &0.4738    &0.3461    &0.0034    &0.0000    &0.0000    &0.0000\\
H-I    &0.0689    &0.0125    &0.0627    &0.0002    &0.0000    &0.0000    &0.0000    &0.0000\\
\bottomrule
\end{tabular}}
\caption{Multiple comparison tests on RFI, DNE and ariaDNE ($\epsilon = 0.02, 0.04,0.06, 0.08, 0.10,0.12 $) values of folivore (Fo), frugivore (Fr), hard-object feeding (H) and insectivore (I). The numbers in the table are $p$ values for the pairwise hypothesis test that the corresponding mean difference is not equal to 0. For $\epsilon = 0.08, 0.10, 0.12$, ariaDNE differentiated folivore, frugivore and hard-object feeding. None of the metrics differentiated insectivore from folivore. }
\label{table:multiplecom}
\end{table}

\begin{figure}
   \begin{center}
   \includegraphics[width=12cm]{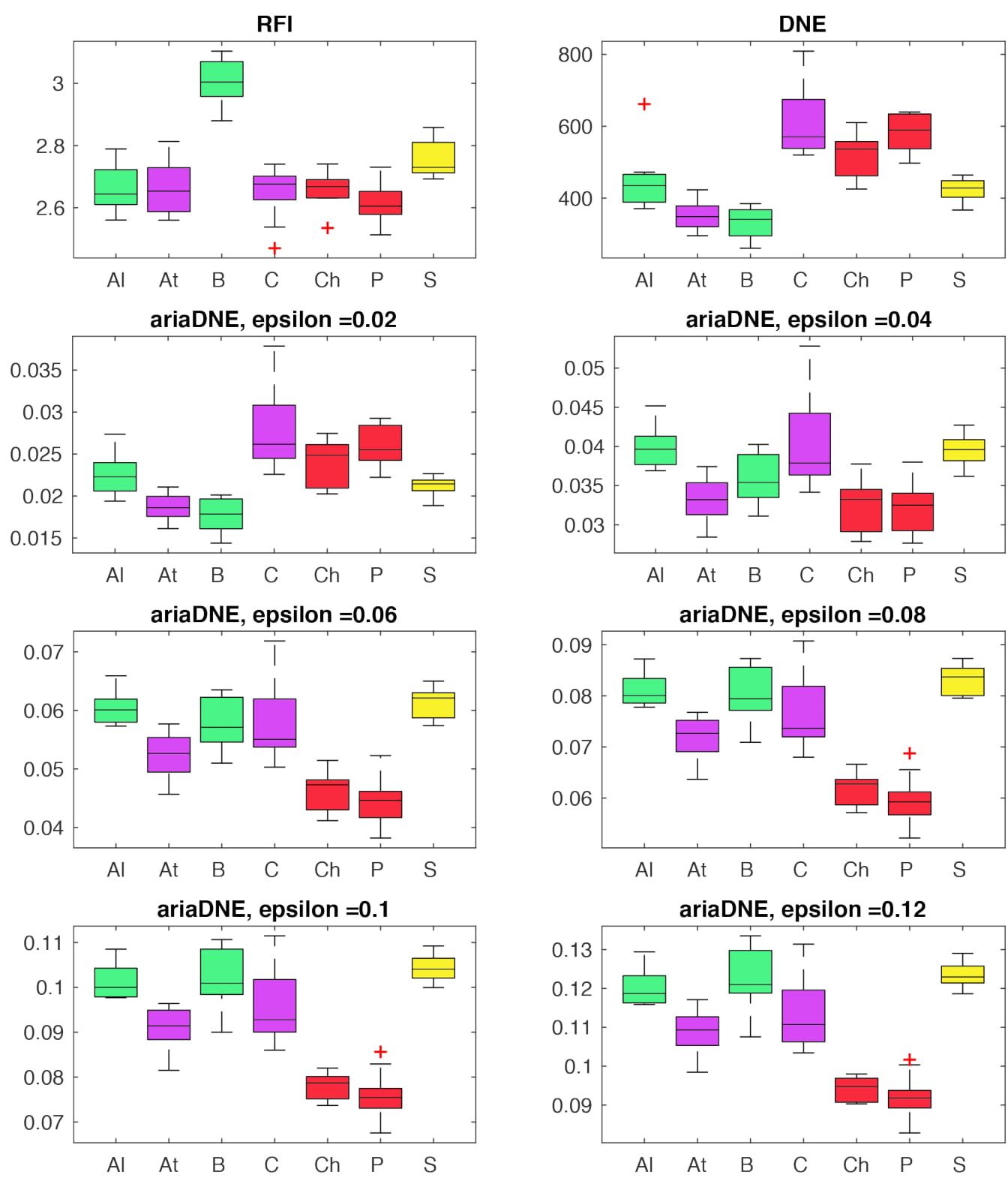} 
    \end{center}
    \caption{Box plots of RFI, DNE, and  ariaDNE with $\epsilon = 0.02, 0.04, 0.06, 0.08, 0.1, 0.12$ for {\it Alouatta}(Al), {\it Ateles}(At), {\it Brachyteles}(B), {\it Callicebus}(C), {\it Chiropotes}(Ch), {\it Pithecia}(P), {\it Saimiri}(S). Color indicates dietary preference: green represents folivore, purple represents frugivore, red represents hard-object feeding and yellow represents insectivore.}
    \label{fig:differentiation}
\end{figure}

For each shape characterizer (RFI, DNE and ariaDNE), ANOVA rejects the hypothesis with $P < 0.05$ that all dietary groups have the same mean, which indicates that some dietary differentiation was detected. To further determine which group means are different, we used multiple comparison tests and the results are summarized in Table~\ref{table:multiplecom}. RFI separated folivore from frugivore and hard-object feeding; DNE in addition separated hard-object feeding from insectivore. As the bandwidth parameter $\epsilon$ increases, ariaDNE further separated frugivore from hard-object feeding and insectivore. No metrics separated folivore and insectivore. However, similarity in their ariaDNE values are not surprising. Insect and leaf tissues tend to be high in structural carbohydrates, which sharpened dental blades are capable of shearing, and therefore high ariaDNE values. What's more important here is the separation from teeth that have low cusps and wide basins, as these are used for crushing motions to efficiently break down soft (i.e., fruit) and hard objects.  For $\epsilon$ =  0.08, 0.1 and 0.12, the box plots of ariaDNE (Fig.~\ref{fig:differentiation}) converge on a pattern in which folivorous {\it Alouatta}, {\it Brachyteles} and insectivorous {\it Saimiri} have higher values, reflecting sharper cusps, whereas frugivorous {\it Ateles} and {\it Callicebus} have lower values and hard-object feeding {\it Chiropotes} and {\it Pithecia} have the lowest values, reflecting low unsharp cusps. The separation was not as clear for RFI and DNE. 

For $\epsilon = 0.02$, ariaDNE shows a pattern similar to DNE. This suggests that when $\epsilon$ is small, both methods capture fine and/or local features on tooth models, and as $\epsilon$ becomes larger, ariaDNE starts capturing larger scale features, ignoring smaller scale features. Figure~\ref{fig:plate} demonstrates the feature scale of DNE and ariaDNE with various $\epsilon$ values. The ariaDNE values for teeth from {\it Callicebus}, {\it Chiropotes} and {\it Pithecia}, which evince less pointed cusps, but which exhibit more fine details on the basin such as enamel crenulations for the Pitheciines, start high when $\epsilon$ is small, but drop with larger $\epsilon$. The pattern is more pronounced in the {\it Pithecia} because their high energy features - the enamel crenulations - are even smaller than those of {\it Callicebus} and so are erased more completely.

It may be hard to assess what is lost or gained by erasing small-scale features. \citeauthor{berthaume2017extant} (\citeyear{berthaume2017extant}) emphasized the importance of small scale features in their analyses of dental topography of extant apes, which also exhibit crenulated enamel similar to Pitheciines. Additionally, erasing small scale features makes the lower second molars of {\it Pithecia} more similar to those of Aye Ayes ({\it Daubentonia}). Previous studies have argued that the two species are analogous from an ecological point of view (\citeauthor{winchester2014dental}, \citeyear{winchester2014dental}).  On the other hand, small scale features could reflect an important functional ability of {\it Pithecia} not available to {\it Daubentonia} (\citeauthor{ledogar2013diet}, \citeyear{ledogar2013diet}). In particular, these small scale features align {\it Pithecia} with {\it Callicebus}, which may be evidence of a close phylogenetic relationship between them - one that was debated prior to availability of genetic data, based on a dearth of obvious unique anatomical similarities. 

\section{Discussion}
\subsection{Bandwidth and multi-scale quantifications}
Even with a less sensitive implementation, ariaDNE still requires choices on the bandwidth parameter $\epsilon$. 
We have discussed the origin and interpretation of $\epsilon$ and how it affects values of ariaDNE in section~\ref{sec:ariadne}, and the resulting differentiation power to dietary preferences in section~\ref{sec:four}. To summarize: (1) for a given $\epsilon$, values of ariaDNE remained relatively unchanged compared to DNE, when the input mesh is perturbed (Fig.~\ref{fig:teaser}). This suggests that $\epsilon$ is independent of mesh attributes like resolution/triangle count, mesh representation, noise level, smoothness, etc. (2) $\epsilon$ indicates the size of local influence: the larger $\epsilon$, the more points on the mesh are considered important to quantify the local energy of the query point, and therefore resulted in larger ariaDNE values.  This means $\epsilon$ determines the scale of features to be included in geometric quantification.  Small $\epsilon$ will make surfaces with finer features have higher ariaDNE values, and large $\epsilon$ will make surfaces with large scale features have higher ariaDNE values. 

Parameter tuning was often achieved through optimization based on a priori goals, yet a single choice of parameter may not satisfy all goals. For example, the parameter that maximizes the differentiation between species in different diet groups may not minimize the effect of wear or optimize the differentiation between species irrespective of diet. 
The requirement of choosing a uniform scale applies to quantitative methods generally, and perhaps this is their biggest weakness compared to qualitative descriptions from more traditional comparative morphology, where multiple scales of perception were naturally integrated. However, the freedom in choosing the parameter also gives possibility in providing more informative comparisons, as seen in (Fig.~\ref{fig:differentiation}). Future work should aim to characterize samples using values computed across a range of $\epsilon$ values.

\subsection{Wider applicability of ariaDNE}
\begin{figure}
 \begin{center}
    \includegraphics[width = 7cm]{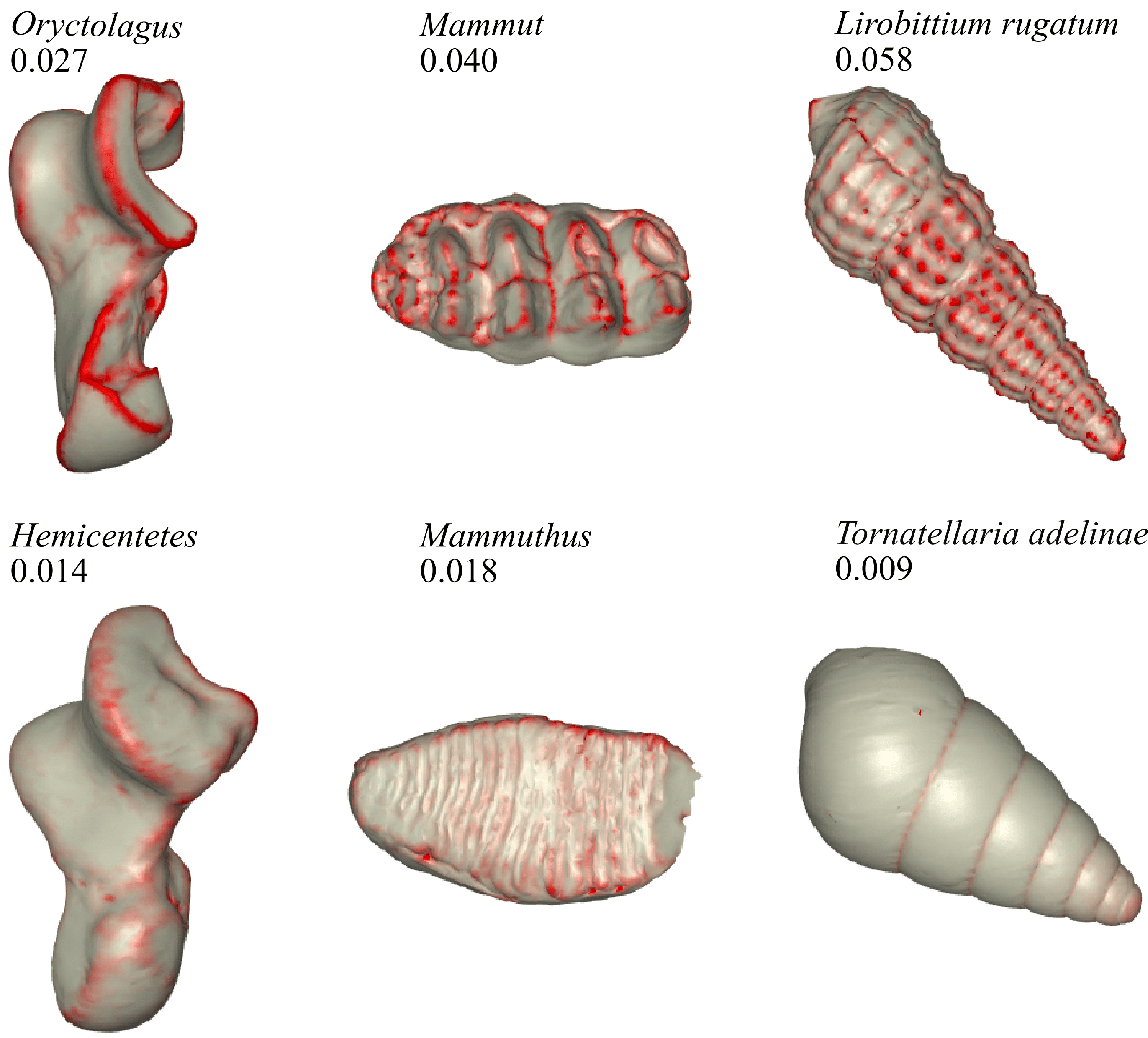}
     \end{center}
\caption{ariaDNE values for surfaces representing astragulus (ankle bone) of {\it Oryctolagus} (saltatorial) and  {\it Hemicentetes} (ambulatory), molars of {\it Mammut} (folivorous) and {\it Mammuthus} (grazing) and shells of {\it Lirobittium rugatum} and {\it Tornatellaria adelinae}. Surface shading indicates curvature computed by our algorithm; ariaDNE values are above each surface. In many cases we expect DNE to covary with mechanical demands of a species environment.}  
\label{fig:examples}
\end{figure}

Many other applications of ariaDNE beyond functional questions of teeth are possible (Fig.~\ref{fig:examples}). For instance, in bivalves, burrowing benthic forms should benefit from shells with greater rugosity (higher ariaDNE) to help them stay embedded in the sea floor, whereas more planktonic forms should benefit from smoother, more hydrodynamic shells (lower ariaDNE). AriaDNE could also be useful for looking at the shape of distal phalanges (bones supporting the nail/claw) as claws suited for climbing are narrower and sharper (higher ariaDNE) while those suited for burrowing (or grasping) will be broader and blunter (lower ariaDNE). In addition, comparing the distribution of ariaDNE values over surfaces will likely provide even more insight into ecologically meaningful shape variation. For example, two surfaces with the same total ariaDNE may have very different distributions: one may have greater spatial variance in ariaDNE, with high ariaDNE features more clustered in one case than another. ariaDNE opens doors to defining other interesting shapes metrics that could potentially assists our understanding in morphology, evolution and ecology.

\subsection{ariaDNE for previously published DNE analysis}
The insensitivity of ariaDNE under varying mesh preparation protocols makes it more widely usable than traditional DNE for comparing and combining results from studies with varying samples or mesh preparation protocols. The computed ariaDNE values for previously published DNE studies (\citeauthor{boyer2008relief}, \citeyear{boyer2008relief}; \citeauthor{bunn2011comparing}, \citeyear{bunn2011comparing}; \citeauthor{winchester2014dental}, \citeyear{winchester2014dental}; \citeauthor{prufrock2016first}, \citeyear{prufrock2016first}; \citeauthor{pampush2016introducing} 2016a, b; \citeauthor{lopez2018dental}, \citeyear{lopez2018dental}) are now available to download as csv files from \url{https://sshanshans.github.io/articles/ariadne.html}. We will continue to update our website as we obtain access to more data samples. 

\subsection{Conclusion}
We provided a robust implementation for DNE by utilizing weighted PCA.  AriaDNE has the advantages of DNE, in that it is landmark-free and independent of initial position, scale, and orientation compared to other popular shape characterizers like OPC and RFI. In addition, ariaDNE is stable under a greater range of mesh preparation protocols, compared to DNE. Specifically, analyses indicated that the new implementation is insensitive to triangle counts, mesh representations, and artifacts on meshes representing both synthetic surface and real tooth data. Additionally, the effects of smoothing and boundary triangles on ariaDNE are limited. AriaDNE retains the potential of DNE for biological studies, illustrated by it effectively differentiating Platyrrhine primate species according to dietary preferences. While increasing the $\epsilon$ parameter of the method can erase small scale features and significantly affect how ariaDNE characterizes structures with small scale features compared to those with larger features (as it did with {\it Chiropotes} and {\it Pithecia} primates in our sample), we think this property can be leveraged to provide more informative comparisons.  Future work should aim to characterize samples using values computed across a range of $\epsilon$ values. In this type of analysis, parameters could be optimized according to model selection criteria. Finally, as with other topographic metrics ariaDNE is likely most informative when deployed in combination with other shape metrics to achieve the goal of more accurately inferring morphological shape attributes.

\section{Acknowledgements}
DMB and JMW were supported in this work by NSF BCS 1552848 and DBI 1661386. ID, SZK and SS were supported in this work by  Math + X Investigators Award 4000837. 

\section{Authors' Contributions}
This research was led by SS. All authors contributed significantly to the manuscript and give final approval for publication. 

\section{Data Accessibility}
\subsection{Sample locations}
The platyrrhine sample we used in this paper was published by \citeauthor{winchester2014dental} (\citeyear{winchester2014dental}), and is available on {\it MorphoSource}, a project-based data archive for 3D morphological data: \url{https://www.morphosource.org/Detail/ProjectDetail/Show/project_id/89}.
\subsection{Matlab scripts} 
\label{sec:code}
Matlab scripts are available from the GitHub repository: \url{https://github.com/sshanshans/ariaDNE_code} and are archived with Zenodo DOI: \url{https://doi.org/10.5281/zenodo.1465949}.

\section{References}
\bibliography{ariaDNE_revision2}
\bibliographystyle{model5-names}\biboptions{authoryear}

\end{document}